\documentclass[runningheads]{svmult}

\usepackage{makeidx}   
\usepackage{graphicx}  
\usepackage{subeqnar}  
\usepackage{multicol}  
\usepackage{physprbb}  
\makeindex             

\begin{document}
\title*{The Unique Signature of Shell Curvature in Gamma-Ray Bursts}
\toctitle{The Unique Signature of Shell Curvature in Gamma-Ray Bursts}
%
%
\titlerunning{}
%
\author{Alicia Margarita Soderberg\inst{1}
\and Edward E. Fenimore\inst{2}}
\authorrunning{Soderberg \& Fenimore}
%
%
\institute{DAMTP, Silver Street, Cambridge CB3 9EW, ENGLAND
\and Los Alamos National Laboratory,
     Los Alamos NM 87545, USA}
\maketitle              
\begin{abstract}
As a result of spherical kinematics, temporal evolution of
received gamma-ray emission should demonstrate signatures of
curvature from the emitting shell.
Specifically, the shape of the pulse decay must bear a strict dependence on the
degree of curvature of the gamma-ray emitting surface.   
We compare the spectral evolution of the decay of individual 
GRB pulses to the evolution as expected from curvature. In particular, 
we examine the relationship between photon flux intensity ($I$) and the
peak of the $\nu F\nu$ distribution
($E_{peak}$) as predicted by colliding shells. 
Kinematics necessitate that $E_{peak}$ demonstrate a power-law relationship 
with $I$ described roughly as: $I=E_{peak}^{(1-\zeta)}$
where $\zeta$ represents a weighted average of the low and high energy spectral
indices.
Data analyses of 24 observed gamma-ray burst pulses provide evidence that there
exists a robust relationship 
between $E_{peak}$ and $I$ in the decay phase.
Simulation results, however, show that a sizable fraction of
observed pulses evolve faster than 
kinematics allow.  Regardless of
kinematic parameters, we found that the existence of curvature demands
that the $I - E_{peak}$ function decay be defined by $\sim (1-\zeta)$.  Efforts were employed to break this
curvature dependency within simulations through a number of scenarios such
as anisotropic emission (jets) with angular dependencies, 
thickness values for the colliding shells, and various cooling
mechanisms.
Of these, the only method successful in dominating 
curvature effects was a slow cooling model.
As a result, GRB models must 
confront the fact that observed pulses do not evolve  
in the manner which curvature demands.
\end{abstract}
\section{Introduction to the Kinematic Model}
The simulated pulses described in this study were created through code based strictly on kinematics.  
\begin{figure}[t]
\begin{center}
\includegraphics[width=.4\textwidth]{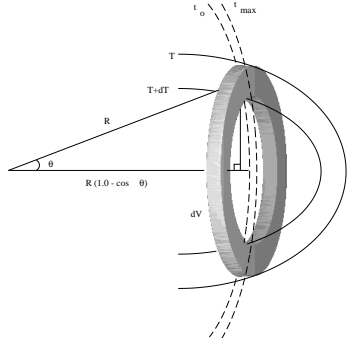}
\end{center}
\caption[]{\scriptsize Geometry of the Kinematic Model}
\normalsize
\label{ring}
\end{figure}
Simulated shells were collided with one another, thereby conserving energy and
momentum and the resulting energy was distributed into 
standard Band function spectra.  Isotropic emission from the merged shell was (initially) assumed where the entirety of the shell is modeled to be gamma-ray active.
Figure 1 demonstrates the geometry of the model.  
 Time of arrival is determined by the angle, $\theta$, at which the emitting patch lies
with respect to the line of sight.  Off axis emission is received
later than on axis emission by a factor of $R(1-cos \theta)$.  
The emitting shell has a slight thickness defined between $R/c=t_{max}$
and $R/c=t_0$.  Photons within shell volume $dV$ contribute to the pulse shape 
between received times, $T$ and $T+dT$.  As a result of the relativistic motion
of the shell, the volume of emitting material which contributes
to the received signal at any time is constant.  Emitting patches on
the shell which fall between the two ellipsoids labeled $T$ and $T+dT$
will arrive at the detector within this range of received time.
\section{Discussion: The Robust Curvature Dependency}
Results of the kinematic studies demonstrate that the 
$I - E_{peak}$ relationship is a robust indicator of shell curvature.  
The strength of this relation was analyzed by varying both Band
and kinematic parameters for the shell model.
Observed pulses, however, do not demonstrate this dependence (see Figures 2 and 3).
\begin{figure}[b]
\begin{center}
\includegraphics[width=.5\textwidth]{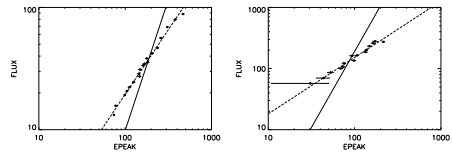}
\end{center}
\caption[]{\scriptsize \bf Comparison with BATSE pulses. \rm Simulations were compared with a data set of 24 pulses selected from the BATSE
GRB Spectral Catalog I (Preece et al.,1999).  
The top figure displays the robustness of the expected and observed
$I - E_{peak}$ relationships for GRB921207.  The solid line represents
the expected decay index as predicted by colliding shells.
It is evident that 
there is a large discrepancy between the data and the simulations.
The same is true for the bottom figure which displays the results for GRB970201.Through these cases, the severity of this discrepancy can be clearly seen.}
\normalsize
\label{fig3}
\end{figure}
In the attempt to break spherical symmetry and reduce the
dependence of pulse shape on curvature effects, 
more complicated emission models were simulated. 
These models
enabled further testing to examine the possibility of
additional dependencies which were not included in the original 
kinematic code.  
Models explored jetting the model emission into
an opening angle between 0.1-5.0 degrees and allowing for intrinsic angular dependencies of the Lorentz factor and/or $E_{peak}$ across the skullcap.
Off-axis shell collisions were simulated such that the collision time
was not instantaneous in the rest frame of the central engine 
and the initial photon emission occurred at an angle outside the 
critical beaming angle.
Various thicknesses were applied to the emitting shell but this only proved to distort the rise time of the pulse and did not have any effect on the shape of the pulse decay or the $I - E_{peak}$ relationship. 
Models also explored fast and slow cooling mechanisms.
Generally, it was found that the curvature dependence is fairly
difficult to break, and requires either grossly distorted geometries 
and/or relatively long cooling time scales.
\begin{figure}[h]
\begin{center}
\includegraphics[width=.35\textwidth]{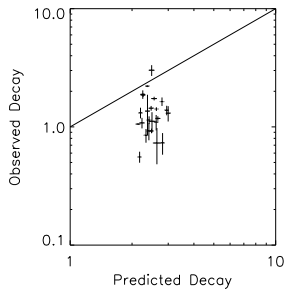}
\end{center}
\caption[]{\scriptsize\bf Comparison of Expected $I - E_{peak}$ Decay with Observed $I - E_{peak}$ Decay from BATSE Pulses. \rm The solid line represents the $I - E_{peak}$ decay index demanded by colliding shells.  Curvature and special relativity impose such a relationship because later portions of the pulse arrive from off-axis emission.  This results
in an expected decay index of $I=E_{peak}^{(1-\zeta)}$.  The majority of BATSE
observations lie below the solid line therefore indicating that observed
$I - E_{peak}$ decay is slower than the decay predicted by kinematics.}
\normalsize
\label{scatter}
\end{figure}
It was found that the emitting shell must cool for
a time period of approximately $t_{cool}=R/c$ in the detector rest frame (where $R/c < \Gamma^2$). This
corresponds to a comoving time of $t'_{cool}=R/{c\Gamma}$ and an
arrival time of $T_{cool}=R/{c\Gamma^2}$ by the standard
transformations.  As a result, the observed cooling time was
comparable in length to the duration due to curvature (e.g. $10^{4}$ s).  Such slow cooling
overwhelmed the curvature dependency with cooling effects throughout
the entire length of the pulse, thereby allowing for a new pulse shape 
evolution.  It is emphasized that slow cooling was the only method 
included in this study which was able to break the robust curvature
dependency as imposed by the kinematics of two colliding shells. \rm 
Typical cooling times, however, are commonly quoted as being shorter than the
duration of the pulse (e.g. $< 10^4$ s).
A remedy to this
situation is to minimize the timescale on which curvature effects can be detected by reducing the radius and/or increasing the bulk Lorentz factor of the emitting shell.   This, in
turn, allows for a relatively shorter
cooling time.  Long cooling times, however, face a number of
problems including efficiency considerations which must be addressed.
%

%

\end{document}